\documentclass[12pt]{article}
\usepackage[latin1]{inputenc}
\usepackage[british]{babel}
\usepackage{cmap}
\usepackage{lmodern}

\usepackage{amssymb, amsmath, amsthm}
\usepackage[a4paper,top=25mm,bottom=25mm,left=25mm,right=25mm]{geometry}
\usepackage{etex}
\usepackage{ragged2e}

\usepackage{authblk} 
\usepackage{pifont}
\usepackage{graphicx}
\usepackage[usenames,dvipsnames,svgnames,table]{xcolor}
\usepackage[figuresright]{rotating}
\usepackage{xtab} 
\usepackage{longtable} 
\usepackage{multirow}
\usepackage{footnote}
\usepackage[stable]{footmisc}
\usepackage{chngpage} 
\usepackage{pdflscape} 
\usepackage[nottoc,notlot,notlof]{tocbibind} 

\usepackage{pgfplots}
\pgfplotsset{compat=1.14}
\usepackage{setspace}

\makesavenoteenv{tabular}
\usepackage{tabularx}
\usepackage{booktabs}
\usepackage{threeparttable}
\usepackage[referable]{threeparttablex} 
\newcolumntype{R}{>{\raggedleft\arraybackslash}X}
\newcolumntype{L}{>{\raggedright\arraybackslash}X}
\newcolumntype{C}{>{\centering\arraybackslash}X}
\newcolumntype{M}[1]{>{\centering\arraybackslash}m{#1}}
\newcolumntype{A}{>{\columncolor{gray!25}}C}
\newcolumntype{a}{>{\columncolor{gray!25}}c}

\newlength{\tablen}

\usepackage{dcolumn} 
\newcolumntype{.}{D{.}{.}{-1}}

\usepackage{tikz}
\usetikzlibrary{arrows, calc, matrix, patterns, positioning, trees}
\usepackage[semicolon]{natbib}
\usepackage[hyphens]{url}
\usepackage{hyperref} 
\hypersetup{
  colorlinks   = true,    
  urlcolor     = blue,    
  linkcolor    = blue,    
  citecolor    = red      
}
\usepackage{microtype}
\usepackage[justification=centerfirst]{caption} 

\usepackage[labelformat=simple]{subcaption}

\DeclareCaptionLabelFormat{parenthesis}{(#2)}
\makeatletter
\renewcommand\p@subfigure{\arabic{figure}.}
\makeatother

\DeclareCaptionLabelFormat{parenthesis}{(#2)}
\makeatletter
\renewcommand\p@subtable{\arabic{table}.}
\makeatother

\usepackage{enumitem}
\setlist[itemize]{leftmargin=2.5\parindent}
\setlist[enumerate]{leftmargin=2.5\parindent}

\theoremstyle{plain}

\theoremstyle{definition}

\theoremstyle{remark}

\def\keywords{\vspace{.5em} 
{\noindent \textit{Keywords}: }}

\def\AMS{\vspace{.5em} 
{\noindent \textbf{\emph{MSC} class}: }}

\def\JEL{\vspace{.5em} 
{\noindent \textbf{\emph{JEL} classification number}: }}

\author{\href{https://sites.google.com/site/laszlocsato87}{L\'aszl\'o Csat\'o}\thanks{~E-mail: \emph{laszlo.csato@uni-corvinus.hu}} }
\affil{Institute for Computer Science and Control, Hungarian Academy of Sciences (MTA SZTAKI) \\
Laboratory on Engineering and Management Intelligence, Research Group of Operations Research and Decision Systems}
\affil{Corvinus University of Budapest (BCE) \\
Department of Operations Research and Actuarial Sciences}
\affil{Budapest, Hungary}
\title{When UEFA rules had inspired \\ unfair behavior on the field}
\date{\today}

\def\Dedication{
{\noindent
``If you can't win, make sure you don't lose.''}\footnote{~Source: They said It: Johan Cruyff, FIFA.com, 25 April 2014. \url{http://www.fifa.com/news/y=2014/m=4/news=they-said-it-johan-cruyff-2323958.html}}
\vspace{0.25cm}

\flushright
\noindent (Johan Cruyff)

\vspace{1cm} 
\justify }

\begin{document}

\maketitle

\Dedication

\begin{abstract}
\noindent
One of the most serious violations of fairness in sports is when a team has clear incentives to lose a match, which hurt a third, innocent team. This is shown to be not only a theoretical possibility as the recently identified incentive incompatibility of the qualification for UEFA club tournaments created such a situation. In particular, we present that SC Heerenveen was ex-ante interested in losing compared to playing a draw on its last match in the 2011/12 Eredivisie, the highest echelon of professional soccer in the Netherlands. In the absence of perverse incentives, the team would probably make more efforts to kick a goal, and a successful attack would send PSV Eindhoven to the more prestigious UEFA Champions League instead of the UEFA Europa League.
Our example may inspire the governing bodies of major sports to consult more with the scientific community, especially before rule changes are implemented.


\AMS{62F07, 91A80, 91B14}

\JEL{C44, D71, Z20}
\end{abstract}

\section{Introduction} \label{Sec1}

Academic researchers pay increasing attention to the issue of fairness in sports. We mention only a few recent works here.
\citet{DuranGuajardoSaure2017} propose a new schedule for the South American qualifiers to the FIFA World Cups, which balances several fairness considerations, and was unanimously approved by the participating countries.
\citet{Vong2017} shows that it is both necessary and sufficient to allow only the top-ranked player to qualify from each group in a multistage tournament to prevent strategic manipulation.
\citet{Guyon2018} scrutinizes the knockout bracket design of the UEFA European Championship 2016, identifies several flaws, and proposes two better procedures.
\citet{BramsIsmail2018} suggest a novel penalty shootout mechanism in soccer in order to reduce the effects of the random coin toss on the outcome.

However, implementation often proves to be difficult in practice. For example, the 133rd Annual Business Meeting (ABM) of The IFAB (International Football Association Board), the rule making body of soccer agreed that the fairer Alternating ($ABBA$) rule \citep{Cohen-ZadaKrumerShapir2018} will no longer be a future option for penalty shootouts due to ``\emph{the absence of strong support}'' \citep{FIFA2018b}. Similarly, FIFA will substantially increase the danger of match fixing with the planned version of the 2026 World Cup \citep{Guyon2019}.

Therefore, we think it is necessary to present a number of clear failures of sports rules in order to persuade administrators and governing bodies on the usefulness of consulting with the academic community.
\citet{KendallLenten2017} offer probably the first comprehensive review of historical sporting rules, which have led to unexpected consequences. Naturally, this work is admittedly not exhaustive: for instance, the authors have not found any match in the history of soccer where a team had clear incentives to lose, and a third team suffered as a result of the unfair behavior. Perhaps the closest case, a match played in the qualifying phase of the 1994 Caribbean Cup between Barbados and Grenada, has not harmed any innocent team, which might have prompted FIFA not to penalize any players \citep[Section~3.9.4.]{KendallLenten2017}.

In the current note, we will give a detailed analysis of a situation when a team was clearly better off by losing compared to playing a draw, and it lost the match, which harmed a third, innocent team. This means our main contribution. 
The example is a better illustration of incentive incompatibility in multiple qualifiers than the hypothetical one presented by \citet{DagaevSonin2018}, which was conditional on a match played later, and finally did not come to pass.

\section{The scandal} \label{Sec2}

The \href{https://en.wikipedia.org/wiki/Eredivisie}{\emph{Eredivisie}} is the top professional league for association football clubs in the Netherlands.
The \href{https://en.wikipedia.org/wiki/KNVB_Cup}{\emph{KNVB Beker}} is a parallel knockout tournament, which is often referred to as the Dutch Cup.
According to the access list for the 2012/13 UEFA club competitions \citep[Annex~IA]{UEFA2012b}, the following teams qualified for the more prestigious UEFA Champions League and the second-tier international competition called UEFA Europa League on the basis of the 2011/12 Eredivisie and KNVB Beker results:
\begin{itemize}
\item
the champion club of the Eredivisie qualified for the group stage of the Champions League;
\item
the runner-up club of the Eredivisie qualified for the third qualifying round of the Champions League;
\item
the third-placed club of the Eredivisie qualified for the play-off round of the Europa League;
\item
the fourth-placed club of the Eredivisie qualified for the third qualifying round of the Europa League;
\item
the fifth-placed club of the Eredivisie qualified for the second qualifying round of the Europa League -- however, the Dutch association chose to organize a special play-off for this place among the teams placed fifth through eighth;
\item
the winner of the KNVB Beker qualified for the play-off round of the Europa League.
\end{itemize}

The cup winner may qualify for the Champions League or the Europa League through the domestic championship, too.
In this case, \citet[Article~2.04]{UEFA2012b} applies:
``\emph{If the winner of the domestic cup qualifies for the UEFA Champions League, the domestic cup runner-up qualifies for the UEFA Europa League at the stage initially reserved for the lowest ranking top domestic league representative. Should both the winner and the runner-up of the domestic cup qualify for the UEFA Champions League, the association concerned may enter for the UEFA Europa League the club which finishes the top domestic league immediately below the other club or clubs which qualify for the UEFA Europa League. In both cases, the access stage initially reserved for the domestic cup winner is reserved for the club which finishes the domestic league in the highest position out of all the clubs which qualify for the UEFA Europa League from the association concerned. Each representative of the domestic league will then enter the competition at the stage initially reserved for the domestic league representative ranked immediately above it.}''

\begin{table}[ht!]
\begin{threeparttable}
\centering
\caption{Ranking of the top teams in the 2011-12 Eredivisie before the last matchday}
\label{Table1}
\rowcolors{1}{gray!20}{}
    \begin{tabularx}{\linewidth}{Cl CCC CCC >{\bfseries}C} \toprule \hiderowcolors
    Pos   & Team  & W     & D     & L     & GF    & GA    & GD    & Pts \\ \midrule \showrowcolors
    1     & Ajax  & 22    & 7     & 4     & 90    & 35    & 55    & 73 \\
    2     & Feyenoord & 20    & 7     & 6     & 67    & 35    & 25    & 67 \\
    3     & PSV Eindhoven & 20    & 6     & 7     & 84    & 46    & 38    & 66 \\
    4     & SC Heerenveen & 18    & 10    & 5     & 77    & 56    & 21    & 64 \\
    5     & AZ Alkmaar & 18    & 8     & 7     & 63    & 35    & 28    & 62 \\
    6     & Twente & 17    & 9     & 7     & 80    & 42    & 38    & 60 \\ \hline    
    \end{tabularx}
    
    \begin{tablenotes}[flushleft]
\item
\footnotesize{Pos = Position; W = Won; D = Drawn; L = Lost; GF = Goals for; GA = Goals against; GD = Goal difference; Pts = Points. All teams have played $33$ matches.}   
    \end{tablenotes}
\end{threeparttable}
\end{table}

Table~\ref{Table1} shows the Eredivisie table before the last matchday played on 6 May 2012. The KNVB Beker was finished on 8 April 2012 when PSV Eindhoven defeated Heracles Almelo by 3-1.

Tie-breaking criteria were: (1) the number of points; (2) goal difference; (3) the number of goals scored. A win was awarded by three points, and a draw by one point.

Ajax would be the champion and Twente could not reach the fourth position independently of their last match. Feyenoord played against SC Heerenveen, PSV Eindhoven against SBV Excelsior, and AZ Alkmaar against FC Groningen on 6 May 2012.

Consider the situation from the perspective of SC Heerenveen:
\begin{itemize}
\item
If it wins, it cannot be placed worse than fourth, so its participation at least in the third qualifying round of the Europa League is assured.
\item
If it plays a draw, it cannot be placed better than fourth. Consequently, it qualifies for the third qualifying round of the Europa League unless both PSV Eindhoven and AZ Alkmaar win. In the latter case, SC Heerenveen is only the fifth (it has a worse goal difference than AZ Alkmaar), furthermore, PSV Eindhoven is the runner-up in the Eredivisie, which creates a vacancy in the Europa League, so Heracles Almelo qualifies for the second qualifying round \citep[Article~2.04]{UEFA2012b}. Then SC Heerenveen should participate with three other teams in a play-off for the place available in the third qualifying round of the Europa League.
\item
If it loses, it cannot be placed worse than fifth. However, Feyenoord will be the runner-up, thus PSV Eindhoven cannot qualify for the Champions League, and SC Heerenveen has a guaranteed place in the third qualifying round of the Europa League.
\end{itemize}

\begin{table}[ht!]
\begin{threeparttable}
\centering
\caption{Scenarios for SC Heerenveen in the 2011/12 Eredivisie before the last matchday}
\label{Table2}
\rowcolors{1}{gray!20}{}
    \begin{tabularx}{1\linewidth}{CCC M{5cm}} \toprule \hiderowcolors
    SC Heerenveen & PSV Eindhoven & AZ Alkmaar & Place in the Europa League \\ \midrule \showrowcolors
	wins  & any result & any result & at least third qualifying round \\ \hline 
    draws & wins  & wins  & play-off for the third qualifying round \\
    draws & does not win & wins  & third qualifying round \\
    draws & any result & does not win & play-off round \\ \hline
    \emph{loses} & \emph{wins}  & \emph{wins} & \emph{third qualifying round} \\
    loses & does not win & wins  & third qualifying round \\
    loses & any result & does not win & play-off round \\ \hline
    \end{tabularx}
    
    \begin{tablenotes}[flushleft]
\item
\footnotesize{The case corresponding to the row in \emph{italics} occurred.}   
    \end{tablenotes}
\end{threeparttable}
\end{table}

Table~\ref{Table2} summarizes all possible cases. It is clear that SC Heerenveen should not play a draw against Feyenoord since losing is an \emph{ex-ante} strictly dominant strategy. In other words, the misaligned UEFA rule may punish this team for its better performance in the national championship.


\section{Assessment} \label{Sec3}

SC Heerenveen lost against Feyenoord by 2-3. Since both PSV Eindhoven and AZ Alkmaar won, SC Heerenveen would be strictly worse off by a draw of 3-3. In the absence of perverse incentives, SC Heerenveen would probably make more efforts to kick a goal in the last minutes, and a successful attack would send PSV Eindhoven to the more prestigious international competition. 

It is a bit mysterious for us but the betting markets seem to be not affected by this event, the odds for the loss (draw) of SC Heerenveen were not especially low (high).\footnote{~See the betting odds, for example, at \url{http://www.betexplorer.com/soccer/netherlands/eredivisie-2011-2012/heerenveen-feyenoord/b3qCaTyl/}, and at \url{http://www.oddsportal.com/soccer/netherlands/eredivisie-2011-2012/heerenveen-feyenoord-b3qCaTyl/}.}

The bizarre situation was recognized by the KNVB (Koninklijke Nederlandse VoetbalBond), the governing body of soccer in the Netherlands as an inherent flaw of the play-off system \citep{KNVB2012}.\footnote{~The Dutch media also discussed this particular case, both before (see \url{https://www.vi.nl/nieuws/heerenveen-mag-in-laatste-wedstrijd-niet-gelijkspelen}) and after (see \url{https://www.frieschdagblad.nl/index.asp?artID=59683}) the match.}
Some commentators thought the cup final should not have played earlier, however, it did not eliminate the problem of incentives. While the absence of the European play-off would have solved this particular problem, the main mistake was committed by the UEFA: according to \citet{DagaevSonin2018}, the only solution guaranteeing incentive compatibility is to award all vacant slots to the team(s) coming from the round-robin tournament, i.e. the Eredivisie.

\section{Conclusions} \label{Sec4}

Tournament organizers supposedly design rules that cannot be manipulated by exerting a lower effort. It never seems to be acceptable if losing dominates winning or playing a draw in soccer.

We have demonstrated that a mistake made in the UEFA Europa League entry rules, identified by \citet{DagaevSonin2018}, probably had led to an unfair behavior in a match played in the 2011/12 Dutch national league, which resulted in a significant financial loss for a third team, and might have upset betting markets. This flaw of the qualification design is corrected from the 2015/16 season in the UEFA Europa League \citep{DagaevSonin2018}, however, the UEFA Champions League qualification has suffered from the same weakness between the 2016/17 and 2018/19 seasons \citep{Csato2019b}. 

Hopefully, the presented example will encourage the governing bodies of major sports to consult more with the scientific community, especially before some rule changes are implemented.

\section*{Acknowledgments}
\addcontentsline{toc}{section}{Acknowledgments}
\noindent
We are grateful to a user (nickname \emph{DatKaiser}) for a comment at \url{https://www.reddit.com/r/soccer/comments/2515jj/question_about_teams_losing_on_purpose/}. \\
We are indebted to the \href{https://en.wikipedia.org/wiki/Wikipedia_community}{Wikipedia community} for contributing to our research by collecting and structuring useful information on the sports tournaments discussed. \\
The research was supported by the OTKA grant K 111797 and by the MTA Premium Postdoctoral Research Program. 

\bibliographystyle{apalike}
\bibliography{All_references}

\begin{thebibliography}{}

\bibitem[Brams and Ismail, 2018]{BramsIsmail2018}
Brams, S.~J. and Ismail, M.~S. (2018).
\newblock Making the rules of sports fairer.
\newblock {\em SIAM Review}, 60(1):181--202.

\bibitem[Cohen-Zada et~al., 2018]{Cohen-ZadaKrumerShapir2018}
Cohen-Zada, D., Krumer, A., and Shapir, O.~M. (2018).
\newblock Testing the effect of serve order in tennis tiebreak.
\newblock {\em Journal of Economic Behavior \& Organization}, 146:106--115.

\bibitem[Csat\'o, 2019]{Csato2019b}
Csat\'o, L. (2019).
\newblock {UEFA} {C}hampions {L}eague entry has not satisfied strategyproofness
  in three seasons.
\newblock {\em Journal of Sports Economics}, in press.
\newblock DOI:
  \href{https://doi.org/10.1177/1527002519833091}{10.1177/1527002519833091}.

\bibitem[Dagaev and Sonin, 2018]{DagaevSonin2018}
Dagaev, D. and Sonin, K. (2018).
\newblock Winning by losing: Incentive incompatibility in multiple qualifiers.
\newblock {\em Journal of Sports Economics}, 19(8):1122--1146.

\bibitem[Dur{\'a}n et~al., 2017]{DuranGuajardoSaure2017}
Dur{\'a}n, G., Guajardo, M., and Saur{\'e}, D. (2017).
\newblock Scheduling the {S}outh {A}merican {Q}ualifiers to the 2018 {FIFA}
  {W}orld {C}up by integer programming.
\newblock {\em European Journal of Operational Research}, 262(3):1109--1115.

\bibitem[FIFA, 2018]{FIFA2018b}
FIFA (2018).
\newblock {IFAB}'s 133rd {A}nnual {B}usiness {M}eeting recommends fine-tuning
  {L}aws for the benefit of the game.
\newblock 22 November 2018.
  \url{https://www.fifa.com/about-fifa/news/y=2018/m=11/news=ifab-s-133rd-annual-business-meeting-recommends-fine-tuning-laws-for-the-benefit.html}.

\bibitem[Guyon, 2018]{Guyon2018}
Guyon, J. (2018).
\newblock What a fairer 24 team {UEFA} {E}uro could look like.
\newblock {\em Journal of Sports Analytics}, 4(4):297--317.

\bibitem[Guyon, 2019]{Guyon2019}
Guyon, J. (2019).
\newblock Will groups of 3 ruin the {W}orld {C}up?
\newblock Manuscript. DOI:
  \href{http://dx.doi.org/10.2139/ssrn.3190779}{10.2139/ssrn.3190779}.

\bibitem[Kendall and Lenten, 2017]{KendallLenten2017}
Kendall, G. and Lenten, L.~J.~A. (2017).
\newblock When sports rules go awry.
\newblock {\em European Journal of Operational Research}, 257(2):377--394.

\bibitem[UEFA, 2012]{UEFA2012b}
UEFA (2012).
\newblock {\em Regulations of the UEFA Europa League 2012-15 Cycle. 2012/13
  Season}.
\newblock
  \url{https://www.uefa.com/MultimediaFiles/Download/Regulations/competitions/Regulations/01/80/06/31/1800631_DOWNLOAD.pdf}.

\bibitem[Voetbalnieuws, 2012]{KNVB2012}
Voetbalnieuws (2012).
\newblock {KNVB} betreurt vreemde situatie voor {SC} {H}eerenveen --
  {F}eyenoord.
\newblock 3 May 2012. \url{http://www.voetbalzone.nl/doc.asp?uid=164548}.

\bibitem[Vong, 2017]{Vong2017}
Vong, A.~I.~K. (2017).
\newblock Strategic manipulation in tournament games.
\newblock {\em Games and Economic Behavior}, 102:562--567.

\end{thebibliography}

\end{document}